# ON DECOMPOSITIONS OF THE KDV 2-SOLITON

NICHOLAS BENES, ALEX KASMAN, AND KEVIN YOUNG

ABSTRACT. The KdV equation is the canonical example of an integrable nonlinear partial differential equation supporting multi-soliton solutions. Seeking to understand the nature of this interaction, we investigate different ways to write the KdV 2-soliton solution as a sum of two or more functions. The paper reviews previous work of this nature and introduces new decompositions with unique features, putting it all in context and in a common notation for ease of comparison.

## 1. INTRODUCTION

The KdV equation is the nonlinear partial differential equation

$$u_t - \frac{3}{2} u u_x - \frac{1}{4} u_{xxx} = 0 \tag{1}$$

for a function $u(x,t)$. Although originally derived over 100 years ago to model surface waves in a canal [14], this simple looking equation has so many interesting features that there is now a category in the Mathematics Classification Scheme (MCS2000) called "KdV-like equations" (35Q53) and has found so many applications in mathematics and physics that it is frequently paired with the adjective "ubiquitous" (see, for example, [8]).

Among its interesting features is the fact that it is completely integrable, and hence that it is possible to write down explicit formulas for many of its solutions. For instance, as was first reported in the 19th century paper by Korteweg and deVries, the equation has a family of 1-soliton solutions

$$u_1(x,t) = u_1(x,t;k,\xi) = 2k^2 \operatorname{sech}^2(\eta(x,t;k,\xi)) \tag{2}$$

$$\eta(x,t;k,\xi) = kx + k^3 t + \xi \tag{3}$$

depending upon the choice of parameters $k$ and $\xi$. Viewing $t$ as a time parameter, these solutions can be described as having a single localized "hump" of height $2k^2$ travelling to the left at speed $k^2$ with position at time $t = 0$ being determined by the value of $\xi$.

It was not until the 1960's that it was recognized that there exist solutions which look asymptotically like linear combinations of two or more of these travelling solitary waves for large $|t|$. Interestingly, although the speeds and heights of the various solitons are the same for $t \to \pm\infty$, the values of the parameter $\xi$ differ, resulting in the famous *phase shift* [27]. (See also [1] where the phase shift is interpreted as a geometric phase in terms of action-angle coordinates under an appropriate Hamiltonian structure.) For instance, in this paper we will be exclusively considering the 2-soliton solution

$$u_2(x,t) = u_2(x,t; k_1, k_2, \xi_1, \xi_2) = 2\partial_x^2 \log(\tau) \tag{4}$$

where here – and liberally throughout the paper – we will make use of the notation

$$\tau = e^{-\eta_1 - \eta_2} + e^{\eta_1 - \eta_2} + e^{\eta_2 - \eta_1} + \epsilon^2 e^{\eta_1 + \eta_2} \tag{5}$$

$$\epsilon = \frac{k_2 - k_1}{k_1 + k_2} \tag{6}$$

$$\eta_i = \eta(x, t; k_i, \xi_i) = k_i x + k_i^3 t + \xi_i \tag{7}$$

and all of the parameters $\xi_i$ and $k_i$ are real numbers such that $0 < k_1 < k_2$.

As shown in Figure 1, for large $|t|$ the solution consists of two solitons moving to the left at speeds $k_1$ and $k_2$ respectively (the illustration shows the case where $k_i = i$ and $\xi_i = \log(3)/2$). That it is not a linear combination of two different 1-solitons is clear from the fact that the maximum height at time $t = 0$ is not the sum of the heights of the peaks at other times. Moreover, the final image which shows the graph of $u_2(x,t)$ over the $xt$-plane makes the phase shift apparent: the nearly linear trajectories of the peaks before and after the collision do not align. (The illustrations in Figure 1 qualitatively represent the generic situation where $k_2$ is much larger than $k_1$. When the difference between them is small, there are *two* local maxima for all time, in contrast to the single maximum shown at $t = 0$ in the figure [16, 17].)

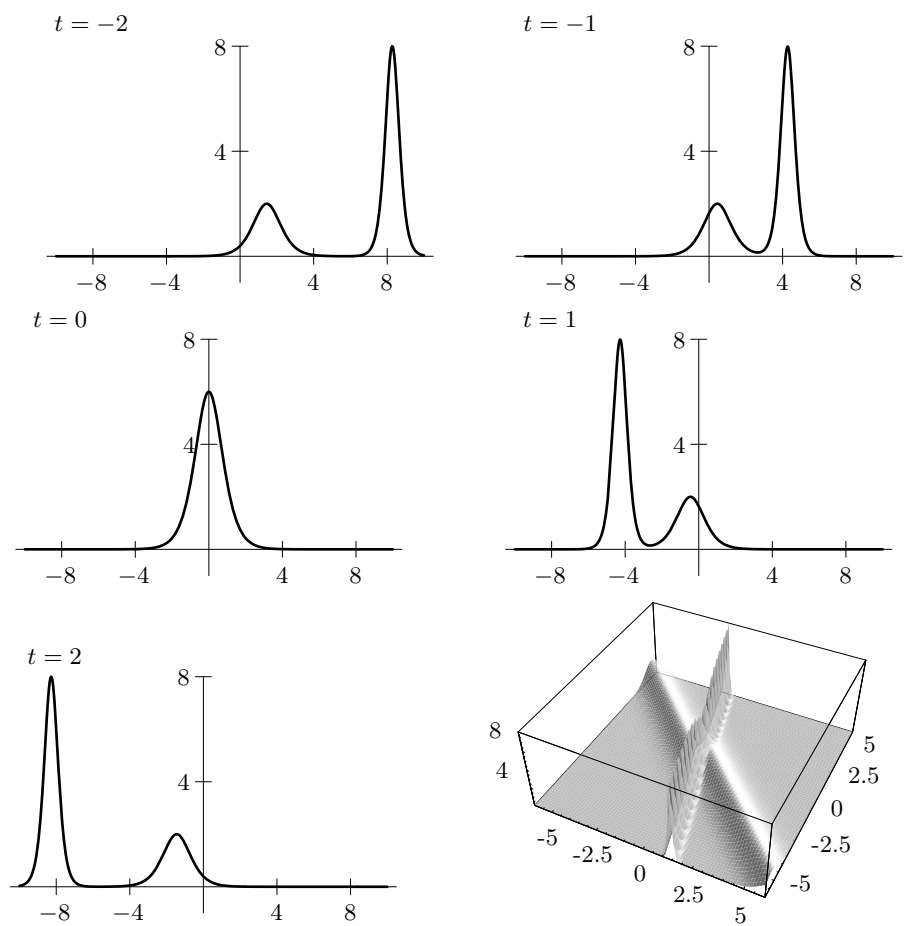

FIGURE 1. A 2-soliton solution of the KdV equation

The standard description of this nonlinear interaction is that the faster soliton is shifted forward while the slower soliton is shifted backwards from where they would have been in a simple linear combination [27]. Note that this description implicitly identifies the peaks before and after the collision based on their speeds. However, there is another possible interpretation: that the rightmost soliton transfers its energy to the leftmost soliton without ever overtaking it. To support this alternative interpretation, we offer the following *decomposition* of the generic 2-soliton solution (4) into a sum of two functions, $u_2(x,t) = f_1(x,t) + f_2(x,t)$:

$$f_1(x,t) = \frac{8\epsilon^2((k_2+k_1)^2 + k_2^2 e^{2\eta_1} + k_1^2 e^{2\eta_2})}{\tau^2} \tag{8}$$

$$f_2(x,t) = \frac{8((k_2-k_1)^2 + k_2^2 e^{-2\eta_1} + k_1^2 e^{-2\eta_2})}{\tau^2}. \tag{9}$$

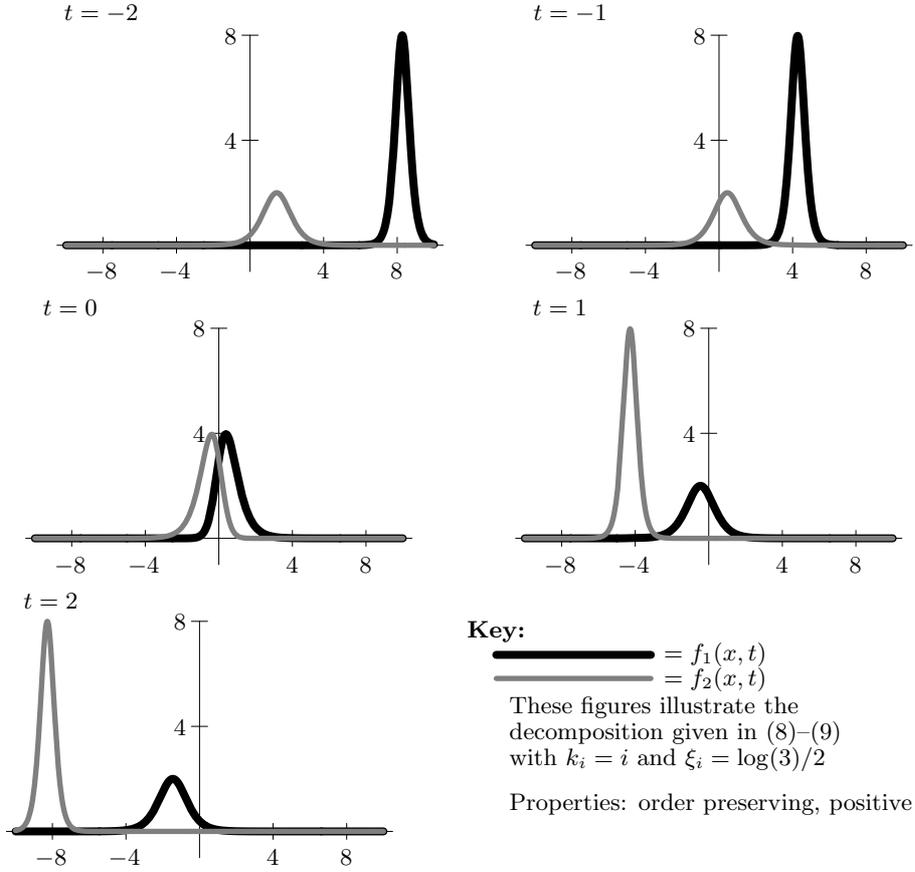

FIGURE 2. The sum of these two functions is the 2-soliton solution shown in Figure 1. It suggests the order preserving interpretation of the KdV 2-soliton in which energy is passed from one peak to the other.

The general case of this decomposition is well represented by the illustrations in Figure 2, in which each of the functions contains one of the two peaks, and they preserve their relative positions but not their speeds. (See Proposition 3.)

Although the decomposition presented in (8)–(9) is new, previous authors have attempted to address this same question by publishing alternative decompositions. Each of the published decompositions has some novel features. However, it has been difficult to compare them because the literature on this subject is scattered and disconnected, because some of the authors provided only existence proofs but no explicit formulae for their decomposition, and because each of the authors uses a slightly different form of the KdV equation (equivalent only up to a change of variables) and their own notation.

It is the goal of this paper to present the first comprehensive survey of previous results on decompositions of the KdV 2-soliton solution, putting the results into perspective, giving a closed formula for each decomposition using common notation, and also to present some novel decompositions which will have not previously appeared in the literature.

2. Asymptotic Decomposition into 1-solitons

We begin our consideration of the two-soliton interaction by examining the asymptotic linear trajectories of the two soliton peaks as $t \to \pm\infty$. Although an analysis of the long-time behavior was done initially in [16], we rederive these results here using the more modern formalism of $\tau$-functions [9].

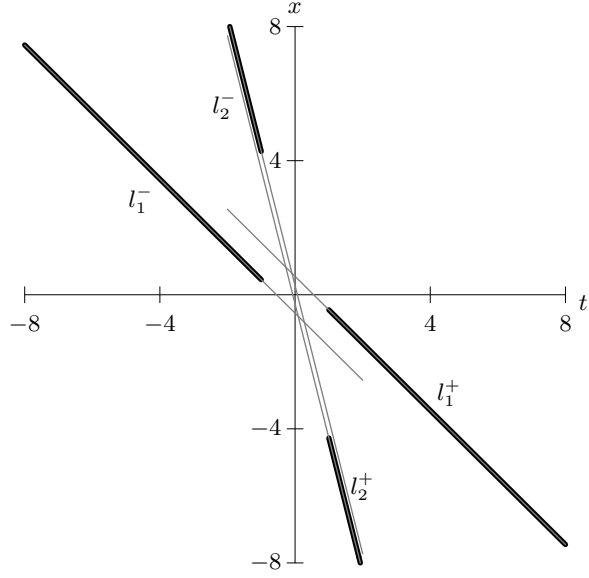

FIGURE 3. The asymptotic trajectories of the peaks in the KdV 2-soliton solution. Note the phase shift which results in two *distinct* lines of each slope.

**Definition 1.** *Let $u_2(x,t)$ be a two-soliton solution to the KdV equation given in (4). We define the following lines:*

$$
\begin{array}{ll}
l_1^- : x = -k_1^2 t - \frac{\xi_1}{k_1} & l_2^- : x = -k_2^2 t - \frac{\xi_2 + \ln \epsilon}{k_2} \\
l_1^+ : x = -k_1^2 t - \frac{\xi_1 + \ln \epsilon}{k_1} & l_2^+ : x = -k_2^2 t - \frac{\xi_2}{k_2}.
\end{array}
\tag{10}
$$

*Also, let $s_i^\pm(x,t)$ denote the following 1-soliton solutions to the KdV equation using the notation from (2)*

$$
\begin{array}{ll}
s_1^- \equiv u_1(x,t; k_1, \xi_1) & s_2^- \equiv u_1(x,t; k_2, \xi_2 + \ln \epsilon) \\
s_1^+ \equiv u_1(x,t; k_1, \xi_1 + \ln \epsilon) & s_2^+ \equiv u_1(x,t; k_2, \xi_2).
\end{array}
\tag{11}
$$

**Proposition 1.** *As $t \to \pm\infty$, $u_2(x,t; k_1, k_2, \xi_1, \xi_2) \to s_1^\pm + s_2^\pm$ which have asymptotic linear trajectories given by the lines $l_1^\pm$ and $l_2^\pm$.*

*Proof.* We first note that since the function $\text{sech}(\eta)$ has a unique local maximum at $\eta = 0$, the peak of the 1-soliton (2) at time $t$ is located at

$$x = -k^2 t - \frac{\xi}{k}. \tag{12}$$

The latter part of the proposition then follows from this fact and Definition 1.

The tau-function formalism is based on the observation that $u_1 = 2\partial_x^2 \log(g(x,t)\tau_1(x,t;k,\xi))$ and $u_2 = 2\partial_x^2 \log(g(x,t)\tau(x,t))$, where $\tau$ is as defined in (5),

$$\tau_1(x,t;k,\xi) = e^{\eta(x,t;k,\xi)} + e^{-\eta(x,t;k,\xi)}, \tag{13}$$

and $g(x,t) = e^{c_1 x + c_2 t + c_3}$ for arbitrary $c_i$. (Multiplication by the function $g$ is what is known as a "gauge transformation" in the tau-function approach to integrable systems since it has no effect on the second logarithmic derivative.)

Then, by writing $\tau$ in terms of $z_i$ and $t$ (where $z_i = x + k_i^2 t$), choosing an appropriate gauge transformation, and taking limits in $t$, we determine the rest of the proposition. Specifically,

$$u_2(x,t) = \lim_{t \to \infty} 2\partial_x^2 \log(e^{-\eta_2 - \log \epsilon}\tau(x,t)) \tag{14}$$

$$= 2\partial_x^2 \log(e^{-\eta_1 - 2\eta_2 - \log \epsilon} + e^{\eta_1 - 2\eta_2 - \log \epsilon} + e^{-\eta_1 - \log \epsilon} + e^{\eta_1 + \log \epsilon}). \tag{15}$$

Let $\chi(z_1, t)$ be the argument of the logarithm above, but written in terms of $z_1 = x + k_1^2 t$ rather than in terms of $x$ and $t$. Using the symbol $\omega$ to denote a function independent of $t$ which is otherwise unimportant, this turns out to be

$$\chi(z_1, t) = e^{-2(k_2^3 - k_1^3)t}\omega(z_1) + e^{-k_1 z_1 - \xi_1 - \log \epsilon} + e^{k_1 z_1 + \xi_1 + \log \epsilon}. \tag{16}$$

The important point is that since $k_1 < k_2$, the first term vanishes as $t \to \infty$ leaving two terms that are independent of $t$ and exactly equal to $\tau_1(x,t;k_1, \xi_1 + \log \epsilon)$. Consequently, in the limit as $t \to \infty$ and in the reference frame moving to the left at speed $k_1^2$ one sees exactly the soliton $s_1^+$. Similarly, using $t \to \infty$ and/or $z_2$ in place of $z_1$ it is agin possible to choose the gauge transformations so that only two terms remain in the limit to prove the rest of the claim. □

In addition to providing another glimpse of the phase shift, the illustration of the four lines $x = l_i^\pm(t)$ in Figure 3 further clarifies the question we seek to address. As time increases (moving to the right) the positions of the peaks move downwards (in the negative $x$-direction) in a nearly linear fashion, except near $t = 0$. There are two ways to identify each of the solitons traveling along these lines for $t \to -\infty$ with one of those as $t \to \infty$. One can imagine a soliton coming in along $l_1^-$ and then at some point being shifted backwards to $l_1^+$ while the other travels along $l_2^-$ until it is shifted ahead to $l_2^+$. Alternatively, this could describe the situation in which a fast soliton comes in along $l_2^-$ until it "bounces" off the other soliton, travelling away along $l_1^+$ while the other soliton similarly was accelerated from its path along $l_1^-$ to $l_2^+$. (See Section 5.1 for more on the implications of this interpretation.)

3. Definitions and Terminology for Decompositions

We say that $\{f_1(x,t), \ldots, f_n(x,t)\}$ is a *decomposition* of the KdV 2-soliton if

(17) $$u_2(x,t) = \sum_{i=1}^n f_i(x,t).$$

Obviously, this definition is very weak. In particular, the functions $f_i$ for $1 \le i \le n-1$ can be chosen arbitrarily and you can still get a decomposition by letting

$$f_n(x,t) = u_2(x,t) - \sum_{i=1}^{n-1} f_i(x,t).$$

There are other properties that such a decomposition can have which would make it interesting:

- We say the decomposition is *positive* if $f_i(x,t) > 0$ for all $(x,t) \in \mathbb{R}^2$. (We also say it is non-negative if $f_i(x,t) \ge 0$.) Note that the decomposition already shown is positive. One nice thing about being positive or non-negative is that the functions in a decomposition do not take any values with large magnitudes where $u_2(x,t)$ is small. (In contrast, some of the decompositions we will see take negative values, which opens up the possibility that they will exhibit visible disturbances away from the two "solitons" of $u_2$.)

- One of the many conservation laws of the KdV equation guarantees that

$$\partial_t \int_{-\infty}^{\infty} u_2(x,t)\, dx = 0.$$

We may similarly want to require such a property for the individual functions $f_i$. So, we say that the decomposition is *mass preserving* if the integral over $\mathbb{R}$ in $x$ of each function $f_i$ is finite and constant for all $t$. (The decomposition already presented is obviously not mass preserving because the functions $f_1$ and $f_2$ have different areas before the interaction and exchange them after.)

- We say that the decomposition is *speed preserving* if for each $i \in \{1, 2\}$ there is a function $f_j$ in the decomposition having a local maximum that travels asymptotically along the path $l_i^-$ for $t \to -\infty$ and along $l_i^+$ for $t \to \infty$ while no $f_j$ has a local maximum travelling along $l_1^-$ in the negative limit and $l_2^+$ in the positive limit. (In other words, these are decompositions which do what the standard description of the soliton interaction says: the solitons preserve their speed but are shifted in the interaction.)

- In contrast, we say that the decomposition is *order preserving* if there is a function $f_j$ in the decomposition which has a local maximum travelling asymptotically along $l_1^-$ and $l_2^+$ in the negative and positive time limits, another function $f_{j'}$ which has a local maximum travelling asymptotically along $l_2^-$ and $l_1^+$ in the negative and positive time limits, but no function in the decomposition that has a local maximum along $l_1^-$ and $l_1^+$ respectively.

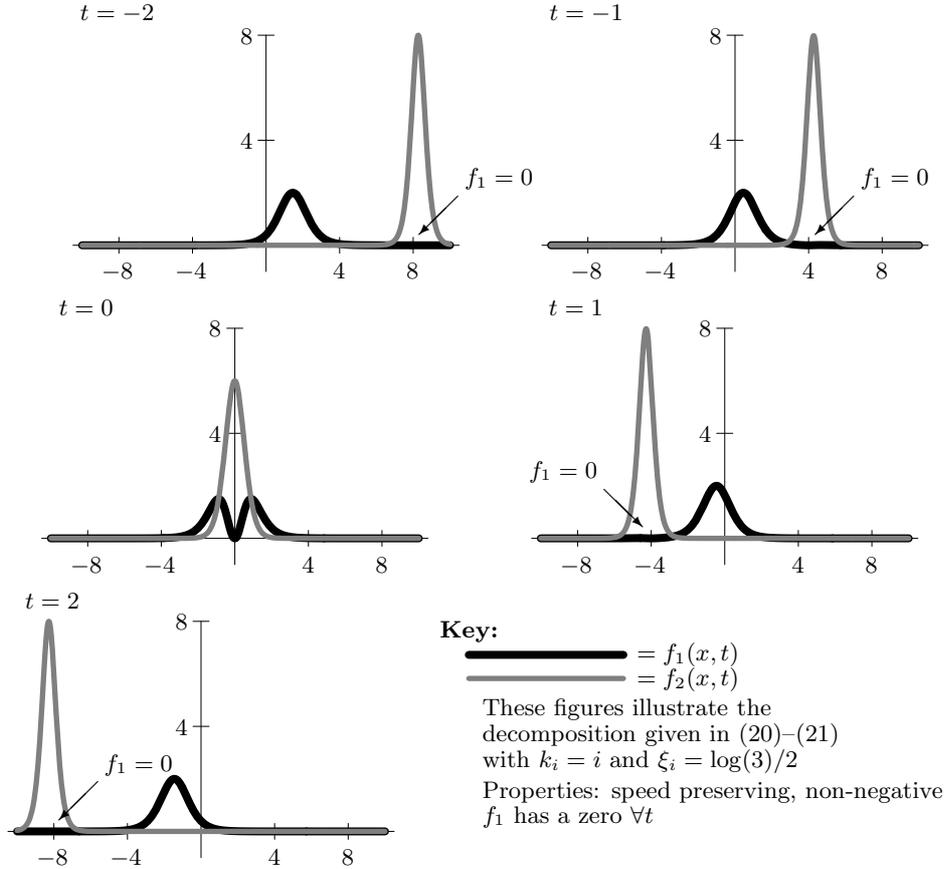

FIGURE 4. The decomposition in (20)–(21) is speed preserving, with the faster soliton overtaking the slower one. Here the phase shift is quite literally given by the individual solitons being shifted forwards and backwards as in the standard description.

It is unfortunate that the last two definitions are somewhat awkward and that not every decomposition can be classified as being either speed preserving or order preserving. However, as the next section will demonstrate, there are many possibilities which must be addressed.

4. A Survey of Decompositions with $n = 2$

4.1. **Speed-Preserving Decomposition of Yoneyama.** In contrast to the decomposition presented in the first section, the oldest published decomposition [5, 10, 26] supports the interpretation of the 2-soliton as a *speed*-preserving interaction. In these decompositions, the faster moving soliton becomes shorter and wider as it overtakes the slower moving one; the slower soliton maintains a zero near the peak of the faster soliton, which gives it the appearance of squeezing its mass underneath as its larger counterpart passes above it. The horizontal stretching of the faster soliton is what makes it shift slightly forward, and the squeezing back of the slower soliton leads to its phase shift. See the illustration in Figure 4.

Yoneyama [26] decomposed the 2-soliton solution in order to better understand the interaction of the solitons and to show that this interaction is *attractive* in nature, *i.e.* that the solitons are pulled toward each other during the interaction. Since an attractive interaction causes the faster soliton to accelerate and the slower to decelerate upon their initial approach, inspection of Figure 3 shows that the decomposition must be speed preserving. Again, this corresponds to the standard description of soliton interaction as described in [27]. Yoneyama also wanted to ensure that his decomposition had a physical justification, so he showed that the functions satisfy the coupled system of equations:

$$(18) \qquad (f_i)_t - \frac{3}{2}u_2(f_i)_x - \frac{1}{4}(f_i)_{xxx} = 0.$$

In these equations, $u_2 \approx f_i$ in the support of $(f_i)_x$ when the solitons are far apart, so these equations approximate the KdV equation and lead to independent soliton behavior. As the solitons approach, they affect each other precisely in the term that makes the equations nonlinear.

In [19], [10] and [5], this same decomposition is reformulated, further investigated and, in the latter, generalized to a broader class of nonlinear evolution equations. The most concise form for the corresponding functions $f_i$ such that $u_2 = f_1 + f_2$ is (cf. [5]):

$$f_i = 2k_i \partial_x (\partial_{\eta_i} \ln \tau). \tag{19}$$

which can more explicitly be written as

$$f_1 = 2k_1 (g(\eta_1, \eta_2))_x \text{sech}^2[g(\eta_1, \eta_2)] \tag{20}$$

$$f_2 = 2k_2 (g(\eta_2, \eta_1))_x \text{sech}^2[g(\eta_2, \eta_1)] \tag{21}$$

$$g(\eta_i, \eta_j) = \eta_i + \frac{1}{2} \ln \left( \frac{1 + \epsilon^2 \exp(2\eta_j)}{1 + \exp(2\eta_j)} \right). \tag{22}$$

In the general case, as in the one illustrated, for large $|t|$ the function $f_i$ looks like a soliton of speed $k_i$, making this decomposition speed preserving rather than order preserving. As noted in [26], this decomposition is mass preserving. However, although it may appear to be positive, it is in fact only *non-negative* since $f_1$ has a zero at $\eta_2 = -\frac{1}{2} \log \epsilon$ (near the peak of $f_2$). Further analysis of this solution was carried out in [6] where it is considered in the context of interacting fields including the so-called "interacton".

4.2. **Mass and Order Preserving Decomposition of Miller and Christiansen.** In [18], Miller and Christiansen acknowledge the problem of soliton identity during collision. The work of [26], [19], and [6] all give mathematical justification for the speed preserving decompositions consistent with attractive soliton interactions. In [2] however, the authors examined the interactions among poles that are seen when soliton solutions to the KdV equation are extended to complex values of $x$. In their investigations, they noted that the poles interact repulsively with the faster moving poles slowing as the slower ones sped up in a manner consistent with an order preserving decomposition. To gain insight into this problem, the authors of [18] develop their own coupled system of equations:

$$(23) \qquad (f_i)_t - \frac{3}{4}\left(u_2(f_i)_x + (u_2)_x f_i\right) - \frac{1}{4}(f_i)_{xxx} = 0.$$

The authors wanted their equations to have the following properties: any solution to them conserves its mass, the equations are symmetric under permutation of indices, they are homogeneous (*i.e.* one can always add components that are identically zero and still satisfy the system), they are linear if $u_2$ is assumed to be a known (non-constant) coefficient, and they are integrable. This last property is demonstrated by placing the coupled equations in the context of the $sl(n+1)$ AKNS hierarchy [18]. Although $u_2$ is only broken into $f_1$ and $f_2$ in [18], the authors note that the above physical properties will be satisfied by a decomposition into any number of solutions as long as the coupled system of equations is satisfied, thus allowing for greater " "degrees of freedom"" than [19] in their decomposition. Furthermore, while mass is conserved in the specific solution that the authors of [19] give to their system of coupled equations it is not conserved for *every* solution; the coupled equations of [18] ensure that mass is conserved for all solutions given the boundary conditions of the $n$-soliton solution, as can be seen from 23.

Since this decomposition is both order preserving (like 8-9) and mass preserving (like 20-21), the functions of the decomposition must take negative values. In particular, $f_2$ starts out including not only the faster moving soliton, but also a region of negative values within the support of $f_1$; this makes the function $f_1$ slightly taller than $s_1^-$. During the interaction, the functions exchange this negative component: the dip of $f_2$ rises and its peak comes down as $f_1$ becomes taller and develops a dip beneath $f_2$. See Figure 5.

Unfortunately, the approach taken in the paper [18] involved existence proofs and numerical simulations only, and no explicit formulas were given for the functions in the decompositions that they studied. We provide the formulas for their decomposition of the KdV 2-soliton $u_2$ in the next proposition.

**Proposition 2.** *The decomposition of Miller and Christiansen is equivalent to*

$$f_1 = \frac{4\epsilon^2 \left( \frac{k_1(k_1+k_2)^2}{k_1-k_2} e^{-2\eta_2} + 2(k_1+k_2)^2 + 2k_2^2 e^{2\eta_1} + k_1(k_1+k_2)e^{2\eta_2} \right)}{\tau^2} \quad (24)$$

$$f_2 = \frac{4 \left( k_1(k_1+k_2)e^{-2\eta_2} + 2k_2^2 e^{-2\eta_1} + 2(k_1-k_2)^2 + \epsilon^2 k_1(k_1-k_2)e^{2\eta_2} \right)}{\tau^2}. \quad (25)$$

*Proof.* Miller and Christiansen provide their solution in terms of the solutions to the linear equation

$$\frac{\sqrt{2}}{3\sqrt{3}} W_t - \frac{1}{\sqrt{6}} \left( \frac{u_2 W}{2} + W_{xx} \right)_x = 0. \quad (26)$$

However, as they point out, these can be found as $W = (\psi(x,t,z)e^{-xz-tz^3})_x$ where $\psi(x,t,z)$ is the Baker-Akhiezer wave function which is an eigenfunction for the operator $\partial_x^2 - u_2(x,t)$ with eigenvalue $z^2$ [25]. Using the method of Darboux transformations to construct the wave function associated to the solution $u_2(x,t)$ [12, 25] we found a closed form for this wave function. According to [18] there should be two values for $z$ which result in solutions $W$ that vanish for $x \to \pm\infty$ and $u_2$ would be their sum. Finding such $z$'s in terms of $k_1$ and $k_2$ resulted in the decomposition above. □

4.3. **Nguyen's "Ghost" Solitons.** Any factorization of $\tau$ gives a corresponding decomposition of the 2-soliton solution $u_2$. Since the tau-function of an $N$-soliton solution is generally computed as the determinant of an $N \times N$ matrix, a natural choice would be the two eigenvalues of the matrix. This is the approach pursued by Nguyen in [21, 22]. (The matrix whose eigenvalues are used is the one related to the dynamics of Ruijsenaars-Schneider particles and which is characterized by rank one conditions [3, 13, 23].) This decomposition is not positive. In fact, as you can see in Figure 6, although the solution looks like two localized, positive peaks before the interaction, one function develops an additional local maximum and the other a corresponding local minimum. Nguyen interprets these as "ghost particles". They persist after the collision and travel faster than $k_2^2$.

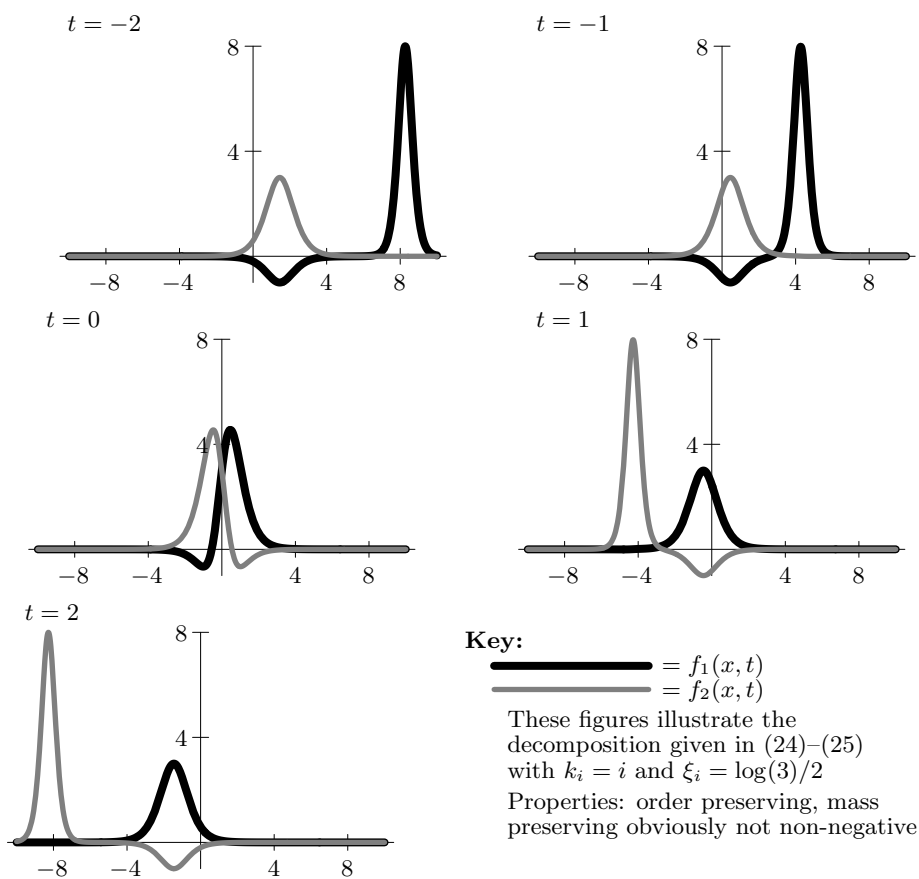

FIGURE 5. This decomposition by Miller and Christiansen was created explicitly to be mass preserving, but since it is also order preserving the functions necessarily take negative values.

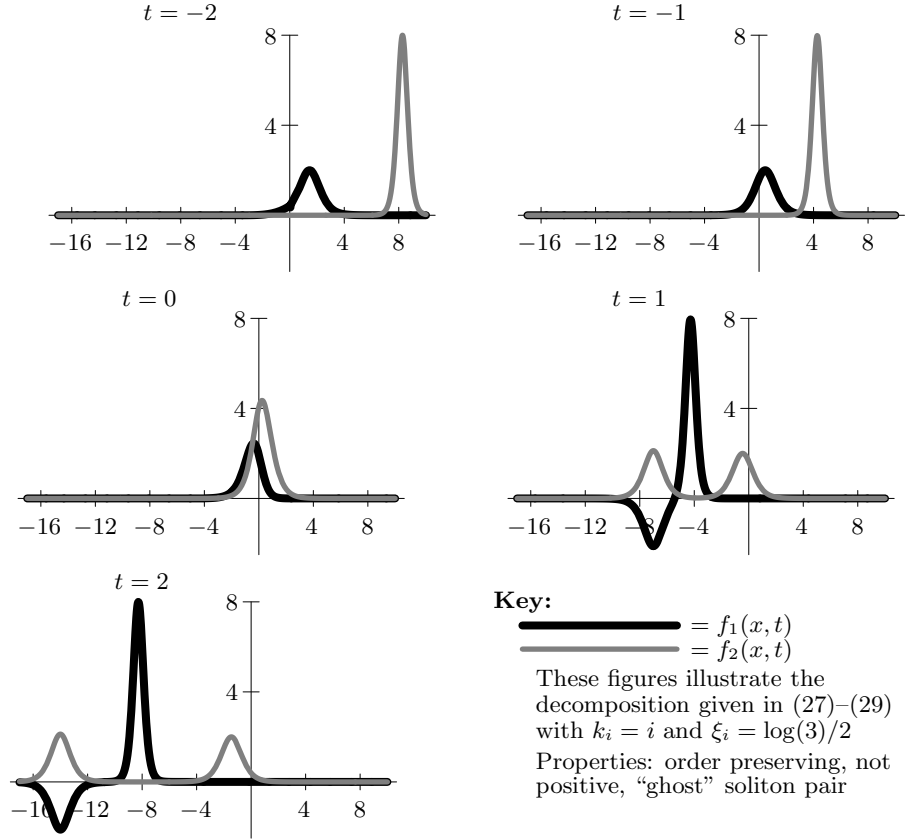

FIGURE 6. Nguyen's decomposition exhibits a "ghost soliton" pair which is produced at the time of the collision. This pair persists and travels off towards $x = -\infty$ faster than either of the solitons.

The functions in this decomposition are

$$f_1 = 2\partial_x^2 \log\left(e^{2\eta_1} + e^{2\eta_2} + 2\epsilon^2 e^{2(\eta_1+\eta_2)} - \sqrt{\gamma}\right) \tag{27}$$

$$f_2 = 2\partial_x^2 \log\left(e^{2\eta_1} + e^{2\eta_2} + 2\epsilon^2 e^{2(\eta_1+\eta_2)} + \sqrt{\gamma}\right) \tag{28}$$

$$\gamma = e^{4\eta_1} + e^{4\eta_2} - \frac{2(k_1^2 - 6k_1k_2 + k_2^2)}{(k_1+k_2)^2} e^{2(\eta_1+\eta_2)}. \tag{29}$$

One unusual feature of this decomposition is that it is not symmetric in time and space. Note that $u_2$ is fixed by an involution which translates and reverses both the $x$ and $t$ axes:

$$u_2(x,t) = u_2(-(x+\gamma_1), -(t+\gamma_2)) \tag{30}$$

where

$$\gamma_1 = \frac{(k_1^3 - k_2^3)\log\epsilon + 2k_1^3\xi_2 - 2k_2^3\xi_1}{k_1^3 k_2 - k_1 k_2^3} \qquad \gamma_2 = \frac{(k_1 - k_2)\log\epsilon + 2k_1\xi_2 - 2k_2\xi_1}{k_2 k_2^3 - k_1^3 k_2} \tag{31}$$

Essentially, this means that you cannot tell if you are watching a 2-soliton running normally or backwards in time and reflected in a mirror. The other decompositions presented in this paper display the same symmetry, either in that each of the functions is preserved under such a transformation or that the functions of the decomposition are exchanged by such a symmetry as in (32) below. However, since the "ghost particles" in Nguyen's decomposition appear after the collision but not before, this decomposition has no such symmetry. (You would know if you were watching it backwards.) Of course, this means that $f_1(-x - \gamma_1, -t - \gamma_2) + f_2(-x - \gamma_1, -t - \gamma_2)$ is *another* decomposition of $u_2$ which is qualitatively different than the one presented by Nguyen; in this case there are ghost particles *prior* to the interaction of the solitons which disappear afterwards.

4.4. **A Novel Decomposition.** We now return to the original decomposition presented in (8)–(9) to explain what properties this decomposition possesses that might generate interest in it. In particular, we need to explain why one might want to consider it as an alternative to the others presented. The answer lies in the simplicity of its formula and its similarity to the soliton solutions of the KdV equation themselves.

Consider that the set of multi-soliton solutions to the KdV equation has the following properties:

- All of its elements are all non-negative, taking only strictly positive values when the parameters and variables are real.
- The set itself is closed under the involution $x \to -x$ and $t \to -t$, which is to say that if one is watching a KdV soliton interaction or the same thing shown in a mirror and run backwards in time. In the case of the 2-soliton solution (4) this symmetry manifests as (30).
- All of its elements take the form of quotients of finite linear combinations of the form $\exp(ax + bt)$.

Note, then, that of the soliton decompositions presented, only ours has all three of these properties. Consequently, we argue that ours is the only decomposition presented thus far in which the component functions are fundamentally like KdV solitons themselves. In particular, (20)–(21) is a decomposition that takes non-negative but never strictly positive values while the other two decompositions involve functions taking negative values, that decompositions (20)–(21) and (27)–(29) necessarily involve square roots linear combinations of exponentials, and that because of the "ghost particles" which only appear after the collision the decomposition (27)–(29) does not reflect symmetry (30).

**Proposition 3.** *The decomposition* (8)–(9) *is positive, order preserving and reflects the symmetry* (30) *through an exchange of the roles of $f_1$ and $f_2$:*

$$(32) \qquad f_1(x,t) = f_2\left(-(x+\gamma_1), -(t+\gamma_2)\right)$$

*where $\gamma_i$ are defined in* (31).

*Proof.* It takes only a simple computation to verify that $u_2 = f_1 + f_2$ and is similarly simple to confirm that the functions take only strictly positive values since $k_i$ and $\xi_i$ are real numbers and everything is then written as a sum or quotient of the squares of such numbers multiplied by exponential functions.

Therefore, all that really requires verification here is the claim that this decomposition has the order preserving property. In particular, we claim that $f_1 \to s_2^-$ and $f_2 \to s_1^-$ as $t \to -\infty$ while $f_1 \to s_1^+$ and $f_2 \to s_2^+$ as $t \to \infty$.

Writing $f_2(x,t)$ instead as a function of $z_1$ and $t$ where $z_1 = x + k_1^2 t$ we find that

$$(33) \quad f_2(z_1, t) = \frac{8\left(k_1^2 + e^{-2k_2(k_2-k_1)t}\omega_1\right)}{\left(e^{k_1 z_1 + \xi_1} + e^{-k_1 z_1 - \xi_1} + e^{-2k_2(k_2-k_1)t}\omega_2\right)^2}$$

where we have used $\omega_i$ to denote terms that are independent of $t$ and will consequently be insignificant. Then, taking the limit as $t$ approaches infinity and recalling that $k_1 < k_2$ we get that

$$(34) \quad \lim_{t \to \infty} f_2(z_1, t) = \frac{8k_1^2}{\left(e^{k_1 z_1 + \xi_1} + e^{-k_1 z_1 - \xi_1}\right)^2} = s_1^-(z_1).$$

A similar argument shows that

$$(35) \quad \lim_{t \to -\infty} f_2(z_2, t) = s_2^+(z_2).$$

That the limits of $f_1$ are also correct can be determined as a consequence of the symmetry (32). $\square$

## 5. Decompositions with $n > 2$

**5.1. Motivations.** Since the KdV 2-soliton looks superficially like a linear combination of two 1-solitons, it seems reasonable to seek decompositions into two functions, as shown above. However, there are reasons one might want to consider decompositions into a sum of three or even four functions.

First, it should be noted that in the original paper by Lax [16] in which the properties of the KdV multi-soliton solutions were first carefully analyzed, there is a discussion of the number of local maxima in the function $u_2$. Regardless of the choice of $k_1$ and $k_2$, for large $|t|$ there are only two local maxima. However, Lax found that near the time of the collision, the number of maxima depends on the ratio $k_2/k_1$. When this ratio is large there will be only one local maximum (as shown in Figure 1), when the ratio is small there will be two for all times, but in a narrow regime in between there are briefly *three* local maxima. This seems to indicate the possibility that there is a third peak that needs to be considered in the decomposition whose existence is normally hidden.

Moreover, there are physical reasons for wanting to consider the case $n > 3$ which grow out of what may at first appear to be a problem for the order preserving interpretation of the soliton interaction. If we are to accept the standard interpretation of KdV soliton collisions, that the faster soliton is shifted ahead and the slower soliton shifted backwards, then we are discussing a situation unlike any physical situation for which we have any intuition. On the other hand, if we identify the solitons before and after the collision by their position, we are describing a familiar situation. It seems quite analogous to the situation in which two billiard balls rolling in the same direction along a straight line meet when the faster ball overtakes the slower ball. In the case of billiard balls, an exchange in energy results from the collision and according to classical physics we would see the trajectories of the balls' centers in the spacetime plane looking like those shown in Figure 7. In this case, the "phase shift" represents nothing other than the sum of the radii of the two billiard balls.

However, there is a problem with this analogy. In the case of the billiard ball collision, the intersection of the in-coming and out-going paths of each of the two balls occur at exactly the same time (indicated by the vertical line in the figure).

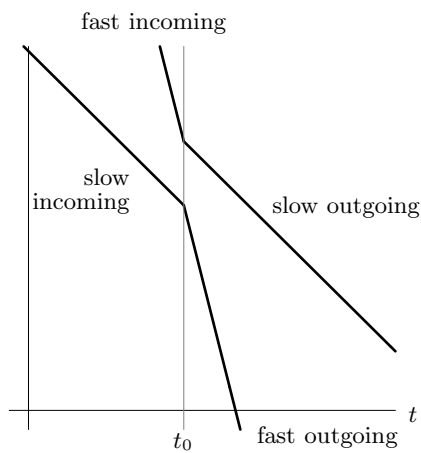

FIGURE 7. An interaction of billiard balls may look superficially similar to the interaction of KdV solitons, but there is an important distinction: the paths intersect at the same time, $t_0$.

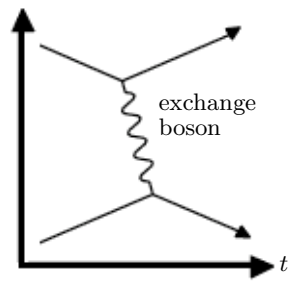

FIGURE 8. This Feynman Diagram of an exchange boson in the interaction of two fermions (essentially copied from [20]) looks more like the KdV soliton interaction in Figure 3.

Yet, this simultanaeity never occurs in the case of KdV solitons. Note that $l_1^-$ and $l_2^+$ intersect at time $t_0$ while $l_2^-$ and $l_1^+$ intersect at time $t_0'$ with

$$(36) \qquad t_0 = \frac{k_1 \xi_2 - k_2 \xi_1}{k_1^3 k_2 - k_1 k_2^3} \;>\; t_0' = t_0 + \frac{\log(\epsilon)}{k_1 k_2 (k_1 + k_2)}.$$

So, if we are to consider the order preserving interpretation of the soliton interaction we have to somehow account for the fact that the faster soliton slows down before the other one speeds up.

One might view this as a reason to reject the notion that the soliton interaction has a physical interpretation as order preserving. But if one looks to quantum physics rather than to the classical dynamics of billiard balls then the interaction of solitons would look quite familiar. The *Feynman Diagram* is a pictoral way to represent the interaction of particles. In these diagrams, the interaction between two particles is not instantaneous but is achieved by the motion of a third particle, an *exchange boson*, as shown in Figure 8. (For more information about these diagrams and to see an image almost exactly like the one reproduced here, see [20].)

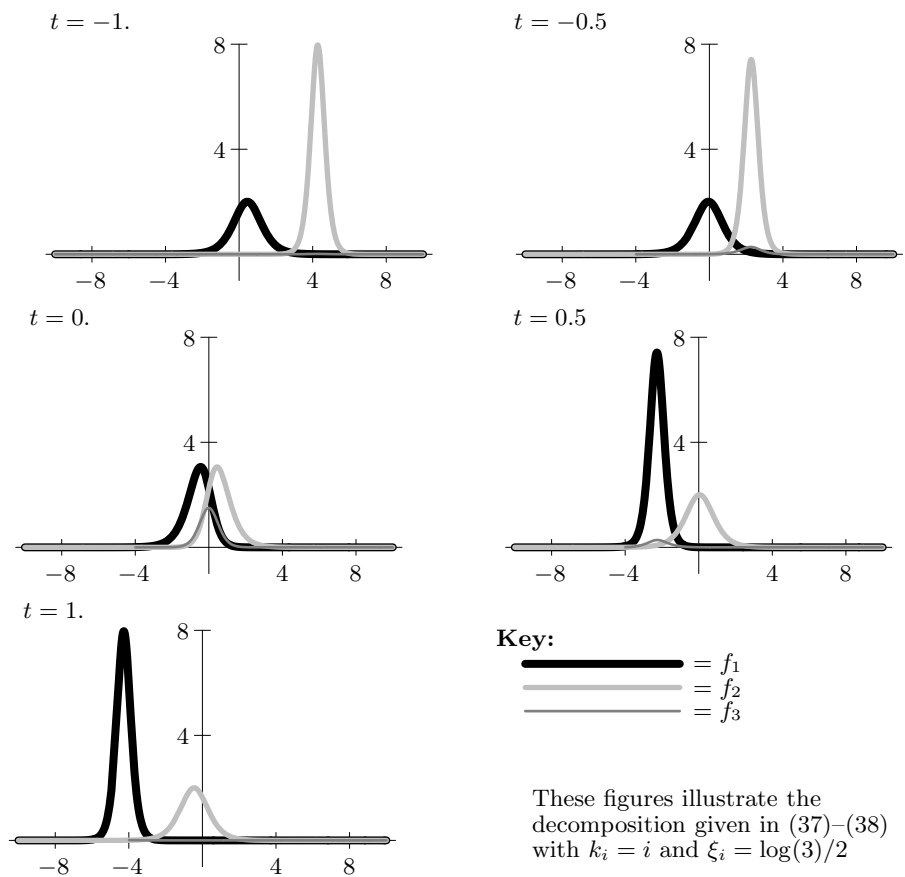

FIGURE 9. The decomposition of Bryan and Stuart exhibiting an "exchange soliton".

It is therefore interesting to note that in the following two sections we present decompositions of $u_2$ into three functions (one from a previous paper by Bryan and Stuart [4] and one new to this paper) that demonstrate a behavior qualitatively like the exchange illustrated in Figure 8.

5.2. **Bryan and Stuart's $n = 3$ decomposition.** The decomposition in [4] seems to be the first to consider the notion of an "exchange soliton". Their decomposition produces exactly what one would hope for in this circumstance: two functions that behave as order preserving solitons and a third function which develops a visible local maximum only during the interaction; see Figure 9.

Their decomposition also starts with the eigenvalues of the same matrix as Nguyen [21, 22] and hence the same function, $\gamma$ from (29), appears in the formulas:

$$f_i = 2\frac{(\mu'_i)^2}{\mu_i(1+\mu_i)^2} \qquad i=1,2 \tag{37}$$

$$f_3 = \sum_{i=1}^{2}(2\partial_x^2 \ln(\mu_i))\frac{\mu_i}{1+\mu_i} \tag{38}$$

where

$$\mu_1 = \frac{(k_1+k_2)e^{-2\eta_1-2\eta_2}}{2(k_2-k_1)^2}\left(e^{2\eta_1}+e^{2\eta_2}-\sqrt{\gamma}\right) \tag{39}$$

$$\mu_2 = \frac{(k_1+k_2)e^{-2\eta_1-2\eta_2}}{2(k_2-k_1)^2}\left(e^{2\eta_1}+e^{2\eta_2}+\sqrt{\gamma}\right). \tag{40}$$

5.3. **An Exchange Soliton Decomposition with a Simpler Formula.** The preceding decomposition is clearly of great interest, although it once again unfortunately requires the introduction of square roots of rational-exponential functions. The new decomposition presented for the first time in this section is qualitatively very much like the one presented in the previous section. However, both the simplicity of its formulae and the path followed by the peak of the "exchange soliton" make it an interesting alternative. It should be viewed as the $n=3$ analogue of our 2 component decomposition (8)–(9).

**Proposition 4.** *The decomposition*

$$f_1(x,t) = \frac{8\epsilon^2(k_2^2 e^{2\eta_1}+k_1^2 e^{2\eta_2})}{\tau^2} \tag{41}$$

$$f_2(x,t) = \frac{8(k_2^2 e^{-2\eta_1}+k_1^2 e^{-2\eta_2})}{\tau^2} \tag{42}$$

$$f_3(x,t) = \frac{16(k_2-k_1)^2}{\tau^2} \tag{43}$$

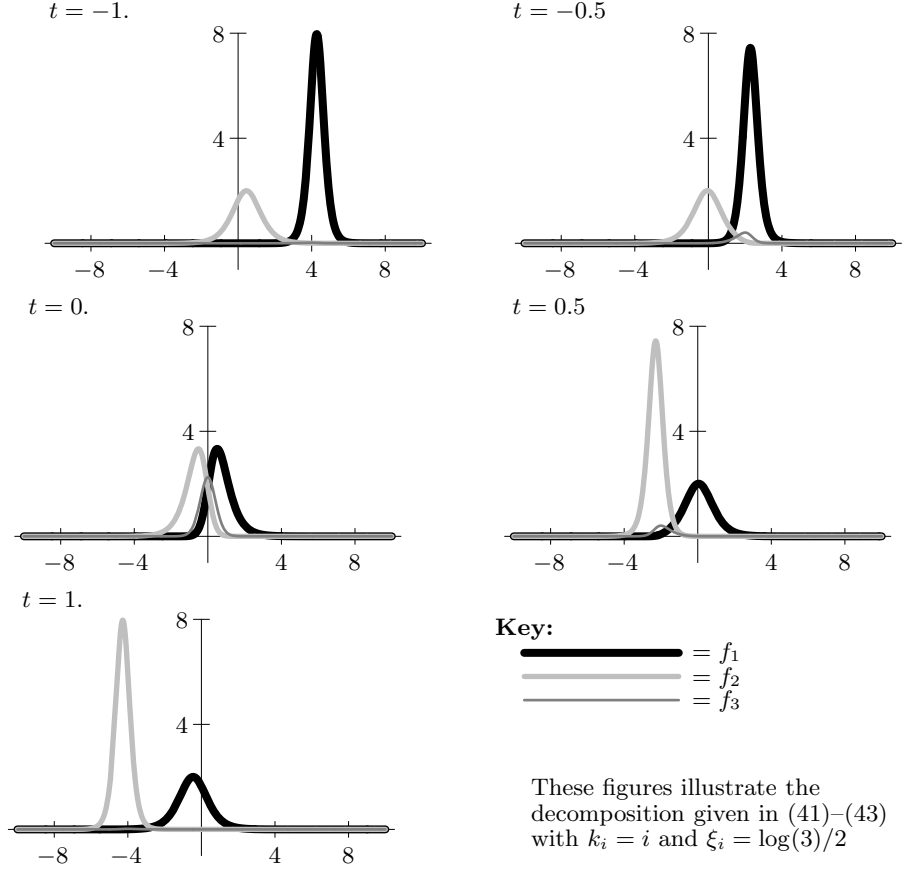

FIGURE 10. A new decomposition with a particularly simple formula exhibiting a "exchange soliton".

*is a positive decomposition into three parts where two are asymptotically order preserving solitons while the third acts as an "exchange" soliton. In particular, the function $f_3$ vanishes for $|t| \to \infty$ and has a unique local maximum for all t located at*

$$\text{(44)} \qquad x = -\frac{1}{k_2}(k_2^3 t + \xi_2 + \log \sqrt{\epsilon}).$$

*Proof.* Since $\lim_{t \to \pm \infty} \tau = 0$ it is clear that the function $f_3$ vanishes as $t$ grows. Then, the simple relationship between this decomposition and the one presented in (8)–(9) provides the necessary asymptotic behavior to conclude that $f_1$ and $f_2$ are again "order preserving solitons". If we let $F_1$ and $F_2$ denote the functions in (8) and (9) respectively then we note that

$$f_1 = F_1 - \frac{f_3}{2} \tag{45}$$

$$f_2 = F_2 - \frac{f_3}{2}. \tag{46}$$

Since the functions all take only positive values and their sum is equal to $u_2$, this is enough to conclude that $f_1$ and $f_2$ have the same order preserving soliton behavior as $F_1$ and $F_2$.

Moreover, we note that $(f_3)_x$ is zero if and only if $x = -(k_2^3 t + \xi_2 + \log(\epsilon)/2)/k_2$. Combined with the fact that $f_3 > 0$ and that $\lim_{x \to \pm\infty} f_3 = 0$ this shows that there is a unique local maximum of $f_3$ which travels along a straightline path in the spacetime plane. □

The easiest way to see that this is not simply the same decomposition as in (37)–(38) written in a nicer form is to compare the height of the function $f_3$ at time $t = 0$ in Figures 9 and 10.

5.4. **Nguyen's $n = 4$ Decomposition.** In [21], Nguyen presents the only published decomposition of a 2-soliton into *four* functions which we know. The motivation is clear: to separate the "ghost" particles visible in the previous decomposition (27)–(29) from the solitons. The result was the decomposition $u_2 = f_1 + f_2 + f_3 + f_4$ where the $f_i$ (written in terms of the functions $\mu_i$ defined in (39)–(40)) are:

$$f_i = 2\frac{\mu_i''}{(1+\mu_i)^2} \qquad i = 1, 2 \tag{47}$$

$$f_i = (2\partial_x^2 \ln \mu_{i-2}) \left(\frac{\mu_{i-2}}{1+\mu_{i-2}}\right)^2 \qquad i = 3, 4 \tag{48}$$

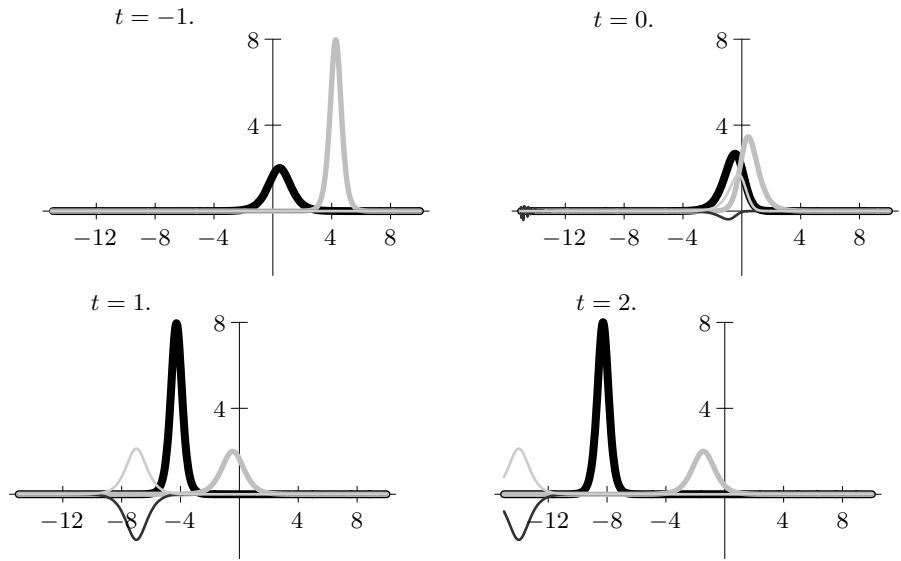

FIGURE 11. A decomposition into *four* functions by Nguyen.

As shown in Figure 11 the graphs of $f_1$ and $f_2$ each contain one of the two peaks of the 2-soliton function $u_2$ while the other two functions develop a local maximum and local minimum that nearly cancel upon addition.

We consider it interesting to note that the function $f_3 + f_4$ behaves qualitatively like the "exchange solitons" of the previous two sections, being positive and becoming large only near the time of the collision. Consequently, $\{f_1, f_2, f_3 + f_4\}$ is yet *another* decomposition of this type.

## 6. Conclusions and Outlook

It should be noted that the question of how to identify the solitons before and after the interactions is not a well posed mathematical problem, and so one should not be expecting a definitive answer. Moreover, there are ways to address the problem other than through decompositions of the form discussed above. In particular, several authors have attempted to provide motivation for the order preserving interpretation by reference to moving "point particles" associated to singularities of solutions of the KdV equation [2, 15, 23].

The wide variety of decompositions reviewed above strongly suggests that there is no unique "best decomposition" in any objective sense. In fact, given any two of these decompositions, it is possible to create another decomposition as a *weighted average* of them. Specifically, if $\{f_i\}$ and $\{g_i\}$ are decompositions of $u_2$ for $1 \leq i \leq n$ (some of the functions can be identically equal to zero if necessary) then so is $\{F(x,t)f_i + (1 - F(x,t))g_i\}$ for an *arbitrary* function $F$. Using such a method to average any of the order preserving decompositions or any of the decompositions exhibiting exchange solitons will result in another decomposition with the same properties. This dramatically demonstrates the extent to which the decompositions fail to be unique.

Perhaps different readers will find some of the decompositions more pleasing or interesting than others. We find it especially interesting to note that the KdV equation is linked to fermions through the construction of the KP hierarchy in terms of particle creation/annihilation operators [11] and to bosons through the interpretation of the tau-function as an example of bosonization [24]. These are the two types of particles displayed in Figure 8. Consequently, we cannot help but wonder whether the similarity between this Feynman Diagram and the solutions displaying an exchange soliton ((37)–(38), (41)–(42) or some suitable average of the two) is more than just a metaphor.

In any case, we believe that the survey of decompositions above is useful in that it provides a variety of valid ways to *think about* the interaction of KdV solitons.

**A**cknowledgements: The first and third authors contributed to the work presented in this paper as part of their studies at the College of Charleston. In particular, for the first author, work on this project constituted an independent study graduate course. The third author was supported financially for his work on this project by the Department of Mathematics as part of an undergraduate research experience. The second author is extremely grateful to these students for their hard work and to the College for this opportunity to work with them.

Department of Mathematics, College of Charleston, Charleston, SC 29424-0001